\documentclass{elsarticle}

\usepackage{amssymb}
\usepackage{amsmath}
\usepackage{float} 
\usepackage{overpic}
\usepackage{pdfpages}

\journal{Urban Climate}

\begin{document}

\begin{frontmatter}



\title{Heatwave-Related Mortality Across Indian Cities Under Future Climate Scenarios}

\author[aff1,aff2]{Ingita Dey Munshi}
\author[aff3]{Abbinav Sankar Kailasam}
\author[aff4,aff5]{Sudeep Shukla}
\author[aff6]{K.~Shuvo Bakar}
\author[aff2,aff5]{Anirban Chakraborti}

\affiliation[aff1]{
             organization={FriskaAi},
             city={Arlington},
             state={Virginia},
             postcode={22203},
             country={USA}
}

\affiliation[aff2]{
             organization={School of Computational \& Integrative Sciences, Jawaharlal Nehru University},
             city={New Delhi},
             postcode={110067},
             country={India}
}

\affiliation[aff3]{
             organization={University College London},
             city={Gower Street, London},
             postcode={WC1E 6BT},
             country={United Kingdom}
}

\affiliation[aff4]{
             organization={AI 4 Water LTD},
             city={Orpington},
             postcode={BR6 9QX},
             country={United Kingdom}
}

\affiliation[aff5]{
             organization={Institute for Interdisciplinary Research},
             city={Chhatarpur, New Delhi},
             postcode={110074},
             country={India}
}

\affiliation[aff6]{
             organization={School of Public Health, University of Sydney},
             city={Sydney},
             postcode={NSW 2050},
             country={Australia}
}

\begin{abstract}
Heatwaves are intensifying as a major climate extreme and have emerged as a growing public health threat in rapidly urbanizing regions such as India. In this study, we integrate long-term heat-related mortality records (1970-2023) with bias-corrected CMIP6 climate projections to quantify future heatwave-related mortality across 67 Indian cities under intermediate (SSP2-4.5) and high-emission (SSP5-8.5) scenarios. A time-series forecasting framework was applied using summer mean temperature as the primary climate driver to project mortality trajectories through the end of the 21st century. Results indicate a strong and sustained increase in heat-related mortality under both scenarios, with multi-fold amplification under SSP5-8.5 relative to SSP2-4.5, reflecting the high sensitivity of health outcomes to emission pathways. Spatial analysis reveals increasing regional divergence under high-emission conditions, with urban regions in the Deccan Plateau, western India, and parts of eastern and northeastern India exhibiting disproportionately higher mortality growth. Multidimensional scaling further highlights emerging clustering of state-level mortality behavior under extreme warming, indicating structurally different regional responses to future heat stress. In contrast, the intermediate mitigation pathway produces more moderate and spatially uniform mortality trends. These findings demonstrate that climate mitigation can substantially reduce both the magnitude and inequality of future urban heat-health burdens. By linking updated climate projections with long-term mortality data at national and sub-national scales, this study provides policy-relevant evidence to support heat adaptation planning and climate-resilient urban development in one of the world’s most heat-vulnerable regions.

\end{abstract}




\end{frontmatter}

\section{Introduction}
\label{sec1}
Heatwaves have emerged as one of the most rapidly intensifying climate extremes under anthropogenic climate change, driven by rising background temperatures and altered atmospheric circulation patterns~\cite{IPCC_2021,Mazdiyasni_2017}. Unlike abrupt disasters such as floods or cyclones, heatwaves often develop gradually and persist over several days, producing cumulative physiological stress and elevated mortality without immediate visual destruction. This characteristic has led to heatwaves being described as “silent killers,” now accounting for an increasing share of weather-related fatalities globally~\cite{Ray_2021,Srivastava_2022}. 

Urban environments are particularly vulnerable to extreme heat exposure. High population density, extensive impervious surfaces, limited vegetation cover, and anthropogenic heat emissions exacerbate temperatures under the urban heat island (UHI) mechanism, above all at night when recovery from accumulated heat stress is crucial for physiology~\citep{Srivastava_2022}.  These urban microclimatic conditions amplify thermal stress during prolonged heat events and increase exposure among vulnerable populations such as the elderly, outdoor workers, and low-income communities with limited access to cooling infrastructure.
India represents a critical hotspot of urban heat vulnerability. Rapid urbanization, combined with high baseline summer temperatures and growing population exposure, has intensified heat-related health risks across metropolitan and secondary cities. Recent assessments indicate that heatwave frequency has increased substantially across India since the mid-20th century, accompanied by rising heatstroke-related mortality and hospitalizations~\citep{Mazdiyasni_2017, Ray_2021}. Urban centers such as Delhi, Ahmedabad, Hyderabad, and cities across the Deccan Plateau and Indo-Gangetic Plain already experience prolonged heat stress during summer months, highlighting the urgent need to quantify future urban heat-health risks under climate change.

Assessing future heatwave-related health impacts requires robust climate projections that capture long-term warming trajectories and associated extremes. The Coupled Model Intercomparison Project Phase 6 (CMIP6) represents the latest generation of coordinated global climate simulations, incorporating improved physical parameterizations and updated forcing scenarios {\citep{Tebaldi_2021}}. CMIP6 introduces Shared Socioeconomic Pathways (SSPs), which integrate greenhouse gas emissions with assumptions about population growth, urbanization patterns, economic development, and technological change.

Compared to earlier Representative Concentration Pathway (RCP) frameworks that primarily focused on radiative forcing trajectories, SSPs provide a more holistic representation of future climate risk by explicitly linking physical climate change with societal development pathways that shape exposure and vulnerability {\citep{Tebaldi_2021}}. For urban heat-risk assessment, this integrated framework is particularly relevant, as future mortality burdens depend not only on temperature increases but also on demographic concentration, urban expansion, and adaptive capacity.
In this study, we analyze a pair of contrasting Shared Socioeconomic Pathway (SSP) scenarios: SSP2-4.5 (an intermediate pathway with moderate climate mitigation and associated warming) and SSP5-8.5 (a high-emission, fossil-fuel-dependent trajectory projecting strong warming by 2100) {\citep{IPCC_2021}}. SSP2-4.5 is widely considered representative of a “current policies” or near-term development trajectory, reflecting emission trends consistent with limited but ongoing mitigation efforts. In contrast, SSP5-8.5 provides an upper-bound warming scenario that captures the potential consequences of continued fossil fuel dependence and weak climate action. Together, these scenarios span a realistic range of plausible futures and enable assessment of how different mitigation pathways may influence long-term urban heat exposure and associated mortality risk across Indian cities. Recent CMIP6-based studies have shown that heatwave frequency is projected to increase substantially across Indian cities, with inland regions exhibiting particularly strong amplification under high-emission scenarios \citep{Rashiq_2026}.

A growing body of literature has documented the increasing health burden associated with extreme heat in India. Mazdiyasni et al. (2017) reported a substantial rise in heatwave frequency and duration since the 1960s, accompanied by strong correlations between sustained high temperatures and mortality {\citep{Mazdiyasni_2017}}. Ray et al. (2021) and Srivastava et al. (2022) further demonstrated that heatwave-related fatalities in India have increased over recent decades, elevating heatwaves to one of the leading causes of weather-related deaths nationwide {\citep{Ray_2021,Srivastava_2022}. Recent CMIP6-based assessments have also highlighted strong regional differences in heatwave behavior across Indian urban environments, emphasizing the importance of city-level analyses for understanding future heat risks \citep{Singh_2024, Rashiq_2026}. Urban heat island effects have been shown to intensify nighttime temperatures by 4-6°C in major Indian cities, exacerbating heat stress and increasing hospitalization and mortality risk {\citep{Srivastava_2022}}.

International studies further highlight the importance of spatial heterogeneity in heat-related mortality. City-level analyses have demonstrated strong variations in vulnerability driven by demographic composition, urban form, socioeconomic conditions, and baseline climate {\citep{Chen_2015, Amoatey_2025, Wang_2024}}.  However, comparable large-scale, long-term assessments for Indian urban regions remain limited.
Several key gaps persist in the current literature. First, many studies focus on short-term associations between temperature and mortality, rather than long-term future trajectories under climate change. Second, earlier projection-based assessments often rely on older CMIP model generations or single-model simulations, limiting robustness and spatial representativeness{\citep{Mazdiyasni_2017}}. Third, national-scale analyses frequently mask sub-national and regional heterogeneity, which is critical for identifying emerging urban heat vulnerability hotspots. Finally, few studies have examined how spatial similarity and divergence in mortality behavior may evolve under contrasting emission pathways, despite its importance for targeted adaptation planning.

To address these gaps, this study integrates multi-model CMIP6 climate projections with long-term heat-related mortality records to quantify future heatwave-related mortality across Indian urban regions under contrasting emission scenarios. The key objectives and contributions of this work are:
\begin{itemize}
  \item To estimate long-term national and state-level trajectories of heatwave-related mortality under SSP2-4.5 and SSP5-8.5 using temperature projections from a six-model CMIP6 ensemble.
  \item To capture spatial heterogeneity in urban heat risk by analyzing 67 geographically diverse Indian cities representing coastal, plain, plateau, and high-altitude regions.
  \item To evaluate regional divergence in mortality behavior under high-emission pathways, identifying emerging spatial clustering patterns using multidimensional scaling analysis.
  \item To provide policy-relevant evidence on how mitigation pathways influence future urban heat-health burdens, supporting targeted heat adaptation strategies and climate-resilient urban planning.
\end{itemize}
By combining updated climate projections, long-term mortality datasets, and spatial pattern analysis, this study provides one of the most comprehensive assessments to date of future heatwave-related mortality across Indian cities under the CMIP6 SSP framework, using city-level climate exposure as a proxy for spatially differentiated urban thermal environments.

\section{Data and methodology}
\subsection{Study area and data}
This study focuses on India’s diverse climatic and geographic regions, spanning coastal zones, arid plains, plateau regions, and high-altitude terrains. To capture spatial heterogeneity in urban heat exposure, 67 representative locations were selected across the country (Fig. 1). These locations correspond to the approximate 1° spatial resolution of CMIP6 climate model grids and include major metropolitan areas as well as secondary urban centers distributed across different climatic regimes.

\begin{figure}[H]
\centering
\includegraphics[width=\textwidth]
{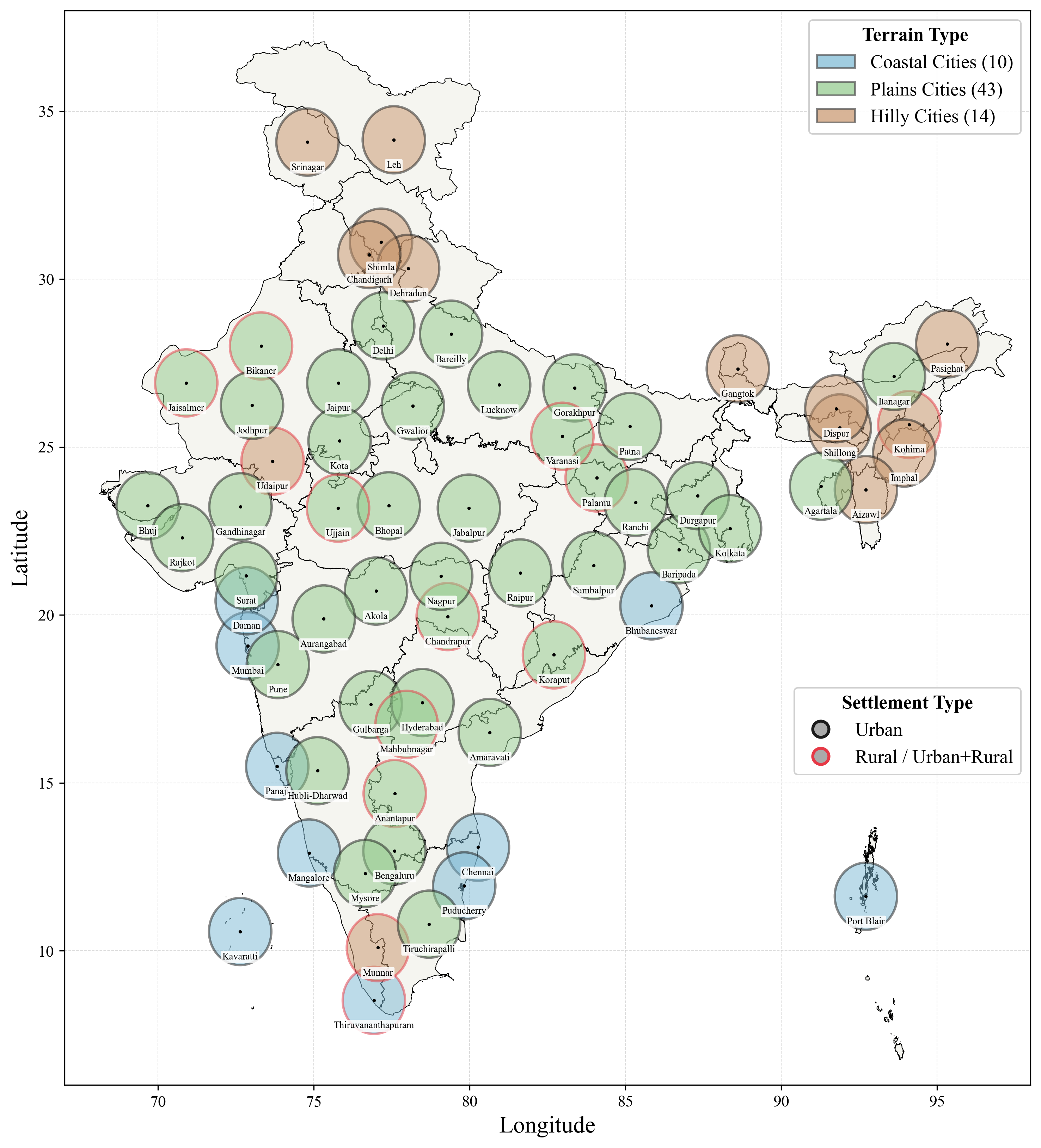}
\caption{Geographic distribution of the 67 urban locations used in this study across India. Selected locations represent diverse climatic regions, including coastal zones, arid and semi-arid plains, plateau regions, and high-altitude environments, and correspond approximately to the spatial resolution of CMIP6 climate model grids.}
\label{fig:locations}
\end{figure}

The selected cities represent coastal, inland plain, plateau, and mountainous environments, enabling analysis of spatial variability in heat exposure and mortality risk across India’s heterogeneous urban landscape. Geographic coordinates of all locations are provided in Supplementary Table S2.

\subsection{Climate Data and Future Scenarios}
Daily maximum temperature (Tmax) and near-surface air temperature at 2 m height (T2M) were obtained from six Coupled Model Intercomparison Project Phase 6 (CMIP6) global climate models: CNRM-CM6-1, CNRM-ESM2-1, EC-Earth3-CC, MIROC-ES2L, MPI-ESM1-2-LR, and MRI-ESM2-0. Model outputs were extracted for the historical period and for two Shared Socioeconomic Pathway (SSP) scenarios: SSP2-4.5 and SSP5-8.5{\citep{Tebaldi_2021}}. These models were selected based on their ability to represent large-scale circulation patterns, land–atmosphere coupling processes, and regional climate dynamics relevant to the Indian subcontinent, as well as the availability of continuous and temporally consistent data across historical and SSP experiments.

To ensure temporal consistency across datasets, all model outputs were standardized to a no-leap calendar format by removing February 29 from leap years. SSP2-4.5 corresponds to a middle-of-the-road pathway featuring moderate mitigation efforts and warming levels, whereas SSP5-8.5 depicts a fossil-fuel-dependent high-emissions pathway projecting pronounced warming by the end of the century {\citep{IPCC_2021}}. An ensemble mean was computed across the six models to reduce individual model biases and improve robustness of projected temperature signals. 
Ensemble averaging was used to emphasize large-scale climate signals and reduce sensitivity to individual model biases, consistent with best practices for multi-model climate impact assessment.

To ensure consistency between climate model outputs and observed temperature magnitudes, a baseline alignment procedure was applied using ground-based observations from the India Meteorological Department (IMD). Summer Mean Temperature (SMT) was calculated for both observed and modeled datasets over the overlapping reference period (1970–2006). The mean offset was computed as the average difference between modeled and observed SMT during this period, and the resulting correction factor was applied uniformly to all projected SMT values for future periods (Fig. S2). This adjustment aligns the modeled temperatures with observations (using 2006 as a representative reference year for scaling) while preserving the modeled temporal trends and reducing systematic bias.

Summer Mean Temperature (SMT) was derived from daily T2M values by averaging over the extended warm season from March to September for each year. Near-surface air temperature (T2M) was selected as the base variable because it directly represents the thermal conditions experienced by human populations and is the standard exposure metric in heat–health epidemiological studies.

The March–September period was chosen following exploratory analysis showing that heat extremes and elevated thermal stress in India frequently extend beyond the conventional April–June summer season (Fig. S1). The analysis indicates that peak annual temperature extremes in India predominantly occur between March and September, with the highest concentration during April, May, and June. However, a non-negligible fraction of extreme temperature events also occurs during the transition periods of the pre-monsoon and early monsoon seasons, particularly in March and July–August for several regions. Based on this combined evidence, the SMT period was defined as March to September to ensure inclusion of the full seasonal window during which extreme heat conditions occur across India. This extended definition captures both early-season and late-season heat exposure, which are increasingly relevant under warming climate conditions.

Although land surface temperature (LST) responds rapidly to radiative forcing, T2M is more closely coupled to human heat exposure through its integration of surface energy balance, atmospheric mixing, and land–atmosphere interactions. SMT therefore serves as a temporally aggregated proxy for cumulative surface thermal conditions affecting populations. At annual and seasonal timescales, cumulative heat exposure metrics such as SMT capture integrated physiological heat stress more effectively than single-day extreme indicators (e.g., daily Tmax). Prolonged exposure to elevated near-surface air temperatures increases cardiovascular and respiratory strain, exacerbates dehydration and heat exhaustion, and reduces nighttime recovery, all of which contribute to elevated heat-related mortality.

SMT exhibited a strong correlation with national-scale heatwave frequency ($r \approx 0.95$) and long-term mortality trends, supporting its suitability as a predictor of population-level heat burden. Previous epidemiological studies have shown that aggregated temperature anomalies provide more stable and robust associations with long-term mortality outcomes than short-duration extremes, particularly when assessing climate change impacts under future scenarios{\citep{Parks_2020}}. Accordingly, SMT was retained as the primary climate predictor for projecting heat-related mortality under different warming pathways.

\subsection{Heatwave Identification Framework}

Heatwave days were identified using the operational thresholds defined by the India Meteorological Department (IMD), which classify heatwave conditions when daily maximum temperature exceeds region-specific absolute thresholds ($\geq 40\,^{\circ}\mathrm{C}$ for plains, $\geq 37\,^{\circ}\mathrm{C}$ for coastal regions, and $\geq 30\,^{\circ}\mathrm{C}$ for hilly areas) {\citep{IMD_2025}}. Percentile-based heatwave definitions {\citep{Cheng_2024, Anderson_2011, Smith_2013}} were also evaluated for sensitivity testing. However, percentile thresholds computed separately for historical and future climate distributions produced artificial discontinuities in heatwave frequency due to climate-driven shifts in the temperature baseline. To preserve temporal consistency across historical and projection periods, the operational IMD threshold definition was retained for the primary analysis.

Although heatwave days were analyzed descriptively, comparative analysis showed strong temporal coherence between heatwave day counts and Summer Mean Temperature (SMT) across both historical and projected periods. Both variables exhibit similar long-term trends and interannual variability, and scatter analysis reveals a strong linear association between SMT and heatwave frequency across emission scenarios ($r \approx 0.95$). Given that SMT provides a continuous measure of cumulative seasonal heat exposure, avoids multicollinearity with discrete heatwave counts, and enhances model stability, it was selected as the primary climate predictor in the mortality forecasting framework.

Annual heat-related mortality data were compiled from official national sources, including the India Meteorological Department (IMD), National Disaster Management Authority (NDMA), Ministry of Statistics and Programme Implementation (MoSPI), Ministry of Earth Sciences (MoES), and the National Crime Records Bureau (NCRB). These datasets provide consolidated records of heatstroke-related fatalities across India from 1970 to 2023. When overlapping records were available for the same year across multiple official sources, an arithmetic mean was computed to derive a consolidated national mortality estimate. This ensemble-averaging approach reduces sensitivity to reporting inconsistencies across agencies and provides a more stable representation of annual mortality variability. Quality control procedures were applied to prioritize official government records and remove inconsistencies across reporting sources.

Long-term city-level mortality datasets are not consistently available across India for the full historical period; therefore, national mortality records were used to estimate the temperature–mortality relationship while spatial heterogeneity was introduced through city-level climate exposure. While mortality records are reported at the national scale, city-level climate exposure was used to capture spatial variability in urban thermal environments across India. This approach enables assessment of regionally differentiated urban heat risk patterns by linking projected temperature variability across urban locations with aggregate mortality trends. Although this framework does not resolve city-specific mortality counts, it allows evaluation of spatial divergence in climate-driven mortality behavior across urbanized regions, which is central to understanding future urban heat vulnerability.

The resulting dataset represents annual aggregated national heat-related mortality counts and was used as the dependent response variable in the ARIMAX forecasting framework.
\subsection{Mortality Forecasting Framework}

Future heat-related mortality projections were generated using an
Autoregressive Integrated Moving Average model with exogenous regressors
(ARIMAX). This time-series modeling framework is well suited for long-term
forecasting when external climate drivers are available and has been widely
applied in temperature–health studies \cite{Masselot_2022,Shah_2024}.
Since mortality data were available at an annual temporal resolution,
no repeating seasonal cycle was present in the time series. Consequently,
a non-seasonal ARIMAX specification was adopted rather than a seasonal
SARIMAX formulation, which is typically used when periodic seasonal
components are present in the data \cite{BoxJenkins1976,Hyndman2018}. The overall methodological workflow integrating climate projections, mortality data, and spatial analysis is summarized in Fig.~\ref{fig:workflow}.

\begin{figure}[H]
\centering
\includegraphics[height=0.97\textheight]{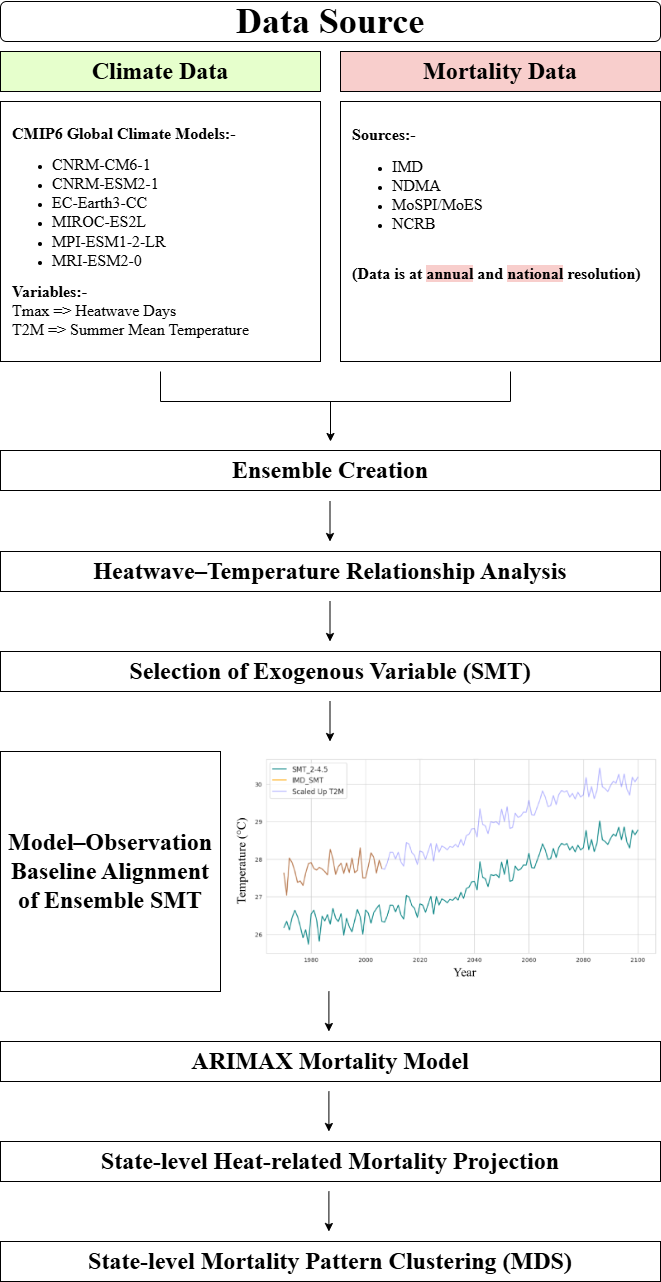}
\caption{Methodological workflow of the heatwave–mortality projection framework.}
\label{fig:workflow}
\end{figure}

ARIMAX was selected in this study due to its ability to capture temporal dependence in mortality time series while maintaining interpretability and numerical stability over long projection horizons. At aggregated annual timescales, temperature–mortality relationships often exhibit approximately linear behavior, particularly when extreme temperature exposure is represented using seasonal or annual mean metrics {\citep{Masselot_2022, Li_2025, Hasan_2025}}. Linear time-series models also offer greater transparency in public health applications compared to complex nonlinear or machine learning approaches, enabling clearer attribution of mortality trends to climate-driven temperature changes.

The primary objective of this modeling framework is to compare long-term mortality trajectories across climate scenarios rather than estimate short-term exposure–response relationships. While distributed lag nonlinear models are commonly used for short-term heat-health attribution, time-series forecasting frameworks such as ARIMAX are better suited for long-horizon scenario analysis where stability, interpretability, and consistency across projection periods are critical.

The ARIMAX framework integrates historical mortality trends with mean-adjusted Summer Mean Temperature (SMT) as the primary exogenous climate driver (Fig. 3). Model parameters were selected to capture annual autocorrelation patterns while avoiding overfitting and preserving model generalizability. The model was trained using historical data from 1970 to 2006 and validated against observed mortality from 2007 to 2023. Separate forecasting experiments were conducted for SSP2-4.5 and SSP5-8.5 scenarios using projected SMT as input to quantify scenario-dependent mortality trajectories.

The ARIMAX model was trained exclusively using national-level annual mortality records and corresponding national mean SMT values. State-level mortality projections were not generated using separate regional model training. Instead, the trained national ARIMAX model was forced with state-aggregated SMT time series to derive relative spatial variations in mortality trajectories while preserving the national mortality scaling. This approach enables consistent spatial differentiation driven by climate exposure patterns without introducing instability associated with training independent regional mortality models.

To assess spatial heterogeneity in future heat-related mortality, projected Summer Mean Temperature (SMT) time series were first aggregated from city-level climate data to administrative state and union territory boundaries. For each state, SMT values were averaged across all corresponding urban locations to generate representative regional climate exposure profiles.

The trained national ARIMAX model was then forced using these state-level SMT inputs to obtain relative regional mortality trajectories. This procedure preserves the national mortality calibration while allowing climate-driven spatial differentiation in projected mortality behavior across regions.

For visualization and spatial pattern analysis, including multidimensional scaling (MDS), projected mortality trajectories were temporally smoothed using an eight-year moving average with a one-year step. This smoothing was applied exclusively for graphical interpretation and was not used during model training or forecasting computations.

\subsection{Spatial Pattern Analysis Using Multidimensional Scaling (MDS)}
To investigate similarities and divergence in state-level heat-related mortality trajectories, multidimensional scaling (MDS) was applied to the projected mortality time series. This dimensionality reduction technique transforms high-dimensional temporal data into a low-dimensional representation while preserving relative dissimilarities between regions. The mortality time series used for MDS analysis represent climate-driven relative state-level projections derived from the nationally trained ARIMAX framework rather than independently calibrated regional mortality models.

MDS was used to identify clustering behavior among states under SSP2-4.5 and SSP5-8.5 scenarios, enabling visualization of spatial convergence and divergence in mortality patterns. State-level mortality trajectories are presented in Fig. 4, while the corresponding two-dimensional MDS embeddings are shown in Fig. 5, allowing direct comparison between temporal trends and spatial similarity structure.

\section{Results}
\subsection{National-Scale Heat-Related Mortality Projections}
The ARIMAX framework captures the historical association between summer mean temperature (SMT) and heat-related mortality and projects a substantial increase in mortality burden across India under both future climate scenarios (Fig. 3). Model validation against observed mortality for the period 2007–2023 demonstrates that the framework reproduces long-term mortality trends with good agreement, supporting its suitability for long-horizon forecasting.

\begin{figure}[H]
\centering
\includegraphics[height=0.55\textheight]{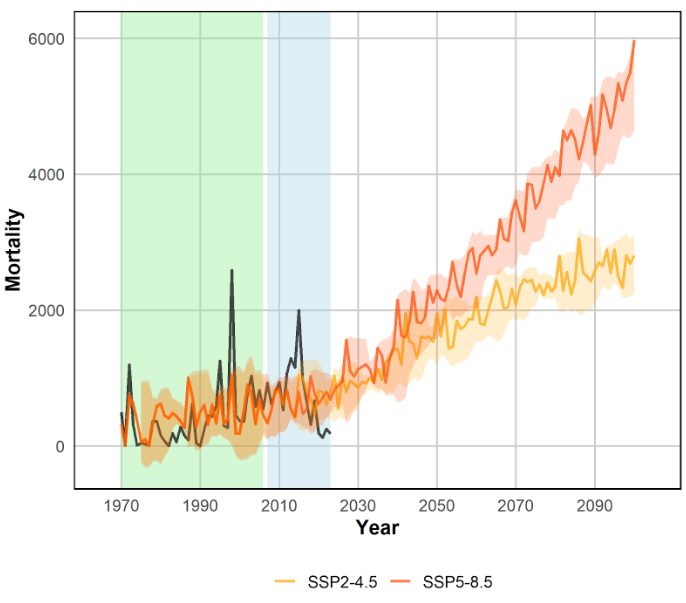}
\caption{Temporal evolution of predicted heat-related mortality from 2024 to 2100 aggregated at the national level. The orange line represents SSP5-8.5 and the yellow line represents SSP2-4.5, showing a steeper mortality increase under the high-emission scenario due to stronger radiative forcing (8.5 W/m$^2$ vs.\ 4.5 W/m$^2$). The model was trained using historical data from 1970 to 2006 and validated using data from 2007 to 2023.}
\label{fig:future_mortality}
\end{figure}

Under the intermediate mitigation pathway (SSP2-4.5), projected mortality exhibits a steady upward trajectory throughout the 21st century, reflecting moderate warming and sustained thermal stress. Unlike the intermediate case, the high-emission SSP5-8.5 pathway leads to a dramatically steeper increase in mortality linked to extreme heat, becoming markedly more accelerated post-mid-century with advancing warming. The divergence between scenarios highlights the strong sensitivity of mortality outcomes to emission pathways and underscores the public health benefits associated with climate mitigation.
These projections indicate that without substantial emission reductions, India is likely to experience a persistent escalation in heat-related mortality, transforming extreme heat from a periodic hazard into a chronic public health burden. The strong divergence between SSP2-4.5 and SSP5-8.5 trajectories highlights the sensitivity of heat-related mortality to emission pathways.

While ensemble-mean projections are presented to highlight central tendencies in mortality trajectories, uncertainty associated with inter-model variability is not explicitly visualized. Differences in climate sensitivity and regional temperature responses across CMIP6 models may influence the magnitude of projected mortality outcomes. These uncertainties should be considered when interpreting absolute mortality estimates, although the relative divergence between emission scenarios remains robust.

\subsection{Spatial Distribution of Future Heat-Related Mortality Across Indian States}
State-level aggregation of projected mortality reveals pronounced spatial heterogeneity in future heat-related risk. Smoothed mortality trajectories for India’s states and union territories under SSP2-4.5 and SSP5-8.5 are shown in Fig. 4.

\begin{figure}[H]
\centering
\begin{overpic}[width=\textwidth]{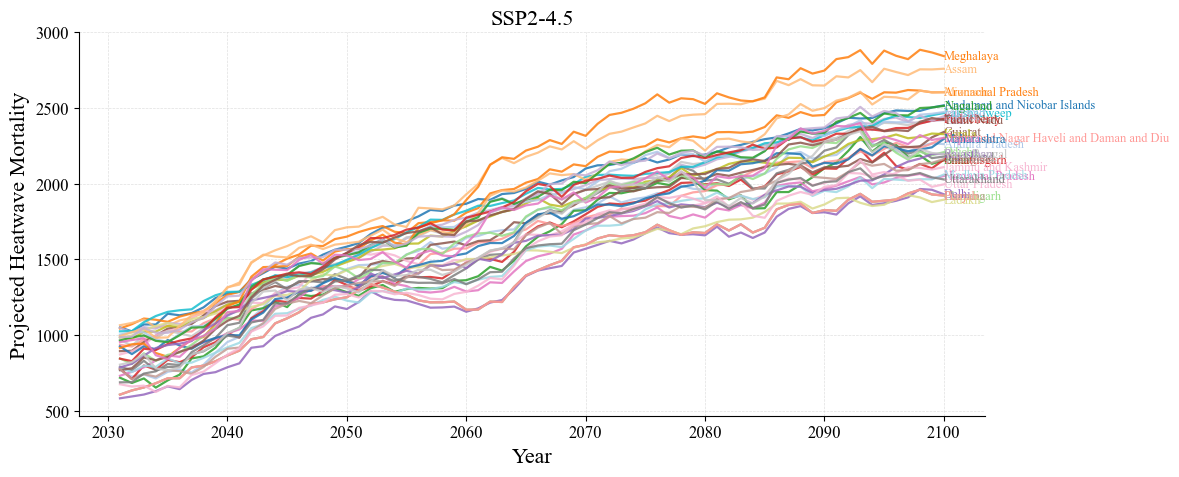}
    \put(1,40){(A)}
\end{overpic}

\begin{overpic}[width=\textwidth]{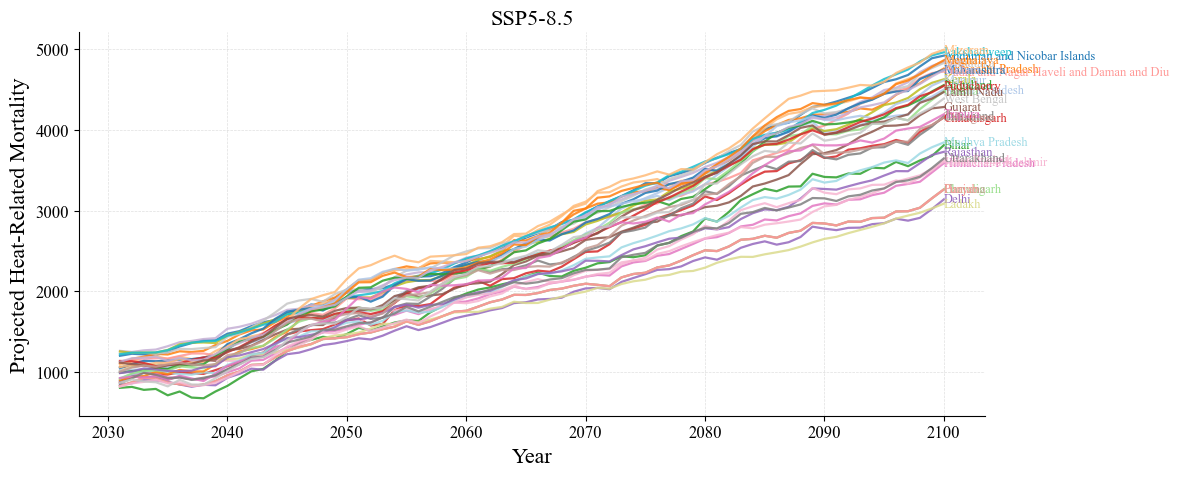}
    \put(1,40){(B)}
\end{overpic}

\caption{State-level projected heat-related mortality trajectories under SSP2-4.5 and SSP5-8.5.}
\label{fig:future_mortality}
\end{figure}

\begin{figure}[H]
\centering
\begin{overpic}[width=\textwidth,height=0.45\textheight,keepaspectratio]{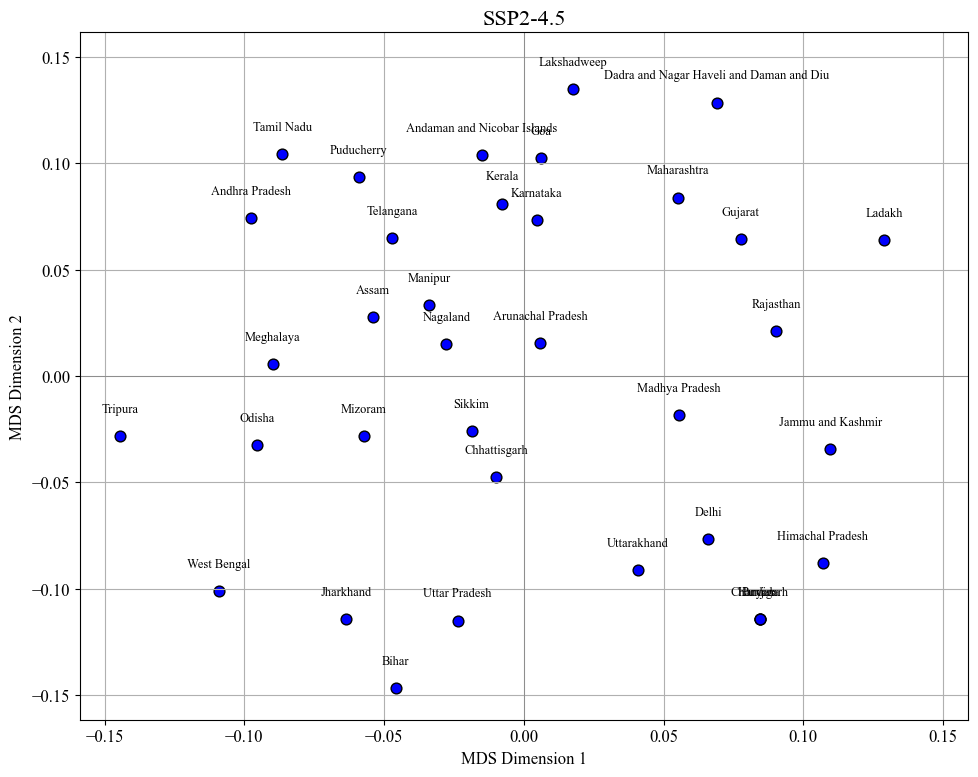}
    \put(1,75){(A)}
\end{overpic}
\begin{overpic}[width=\textwidth,height=0.5\textheight,keepaspectratio]{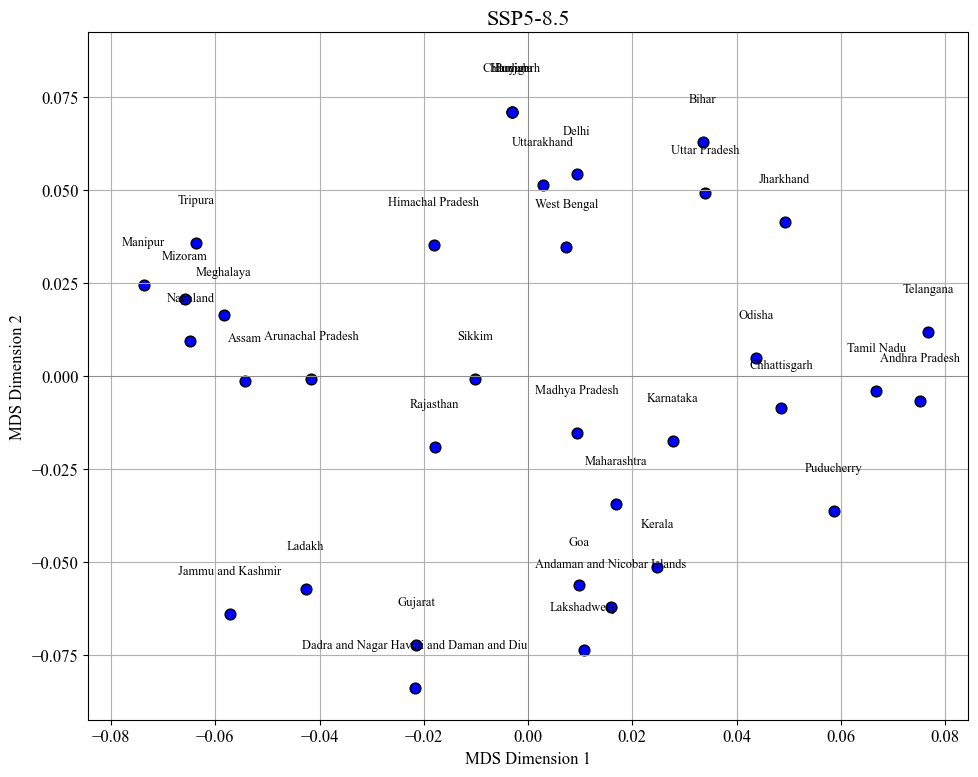}
    \put(1,75){(B)}
\end{overpic}
\caption{Two-dimensional multidimensional scaling (MDS) representation of state-level mortality trajectories showing spatial similarity patterns.}
\label{fig:future_mortality}
\end{figure}

Under SSP2-4.5, most states display broadly similar upward trends (Fig. 4A), suggesting relatively uniform growth in heat-related mortality at moderate warming levels. However, under SSP5-8.5, regional disparities become increasingly apparent (Fig. 4B). States located in the Deccan Plateau, northwestern arid regions, and parts of eastern and northeastern India exhibit more rapid mortality increases compared to cooler or high-altitude regions.

These spatial patterns reflect the combined influence of baseline climate, projected temperature amplification, and regional exposure characteristics. Hot semi-arid and tropical regions, already operating near physiological heat stress thresholds, show amplified sensitivity to additional warming, while high-altitude regions exhibit comparatively slower mortality growth despite warming trends.

\subsection{Regional Divergence and Clustering of Mortality Trajectories}
Under the intermediate emission scenario, state-level mortality trajectories remain relatively clustered, with limited divergence across regions (Fig. 5A). This pattern indicates that moderate warming produces broadly consistent mortality responses across India, with differences primarily driven by baseline temperature gradients rather than strong nonlinear amplification.
The relative uniformity of trajectories under SSP2-4.5 suggests that mitigation-oriented pathways can limit the emergence of extreme spatial inequality in urban heat-related mortality risk.

In contrast, SSP5-8.5 produces pronounced divergence among state-level mortality trajectories (Fig. 5B). Several regions exhibit accelerated mortality growth, while others follow comparatively flatter trajectories. This widening separation reflects the nonlinear amplification of heat stress under extreme warming conditions.

Hot climate regions such as western and central India, along with densely populated plateau states, show the most rapid mortality escalation. Meanwhile, Himalayan and high-altitude regions display slower growth despite warming, highlighting the role of baseline climate and physiological thresholds in shaping regional sensitivity

These results indicate that high-emission futures not only increase overall mortality burden but also intensify spatial inequality in heat-related health impacts. This contrast becomes particularly evident by the end of the century (Fig. 6), where the 2100 spatial distribution reveals markedly stronger regional concentration of mortality under SSP5-8.5 compared to the more moderate and relatively uniform pattern under SSP2-4.5.

\begin{figure}[H]
\centering
\includegraphics[width=\textwidth]
{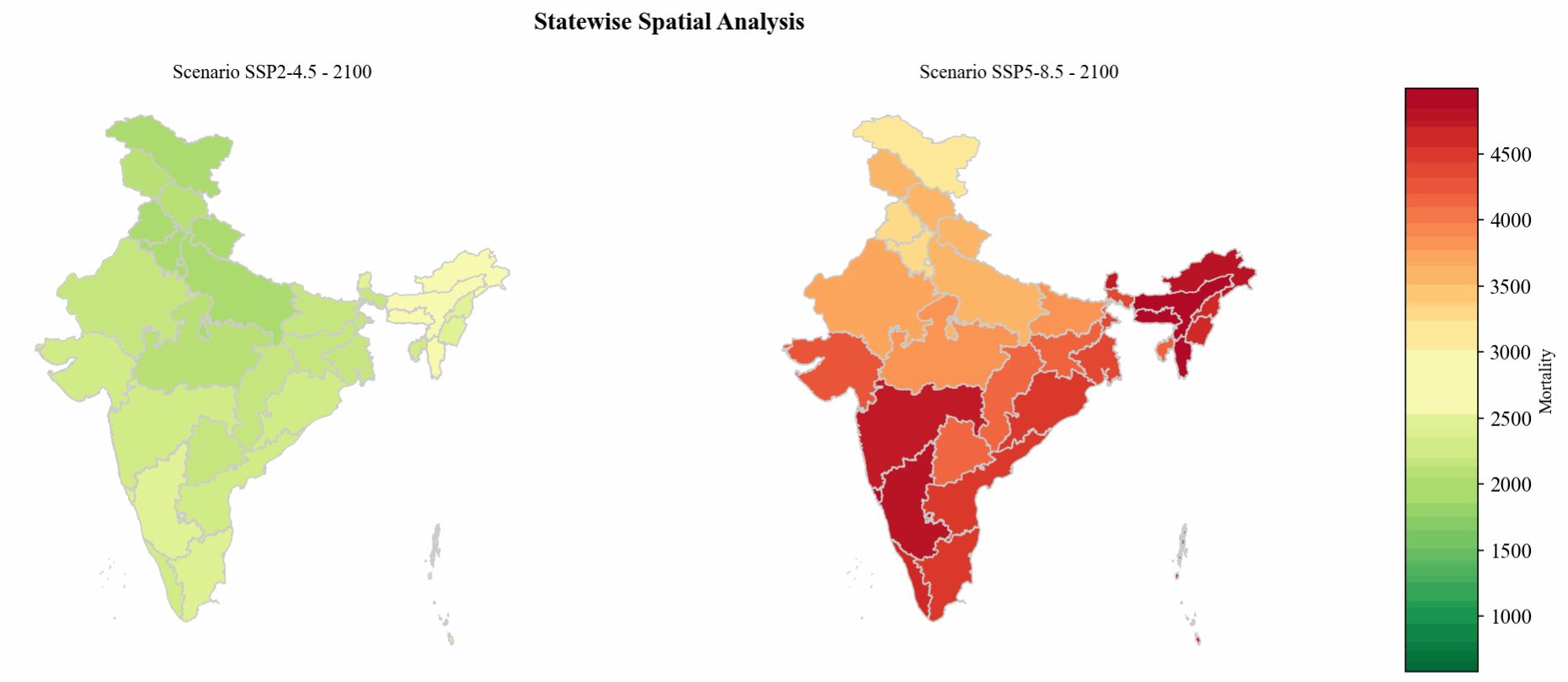}
\caption{State-level distribution of projected heat-related mortality in 2100 under SSP2-4.5 (left) and SSP5-8.5 (right). The high-emission pathway exhibits pronounced regional concentration and widening spatial inequality compared to the more moderate and relatively uniform distribution under SSP2-4.5.}
\label{fig:locations}
\end{figure}

An animated visualization of the full temporal evolution of state-level mortality patterns under both scenarios is provided in the Supplementary Material (Video S1).

\subsection{Multidimensional Scaling Analysis of State-Level Mortality Behavior}
To further examine spatial similarity patterns, multidimensional scaling (MDS) was applied to state-level mortality time series. Under SSP2-4.5, the two-dimensional embedding shows tight clustering of states (Fig. 5A), confirming the relatively homogeneous mortality behavior observed in the time series analysis.
Under SSP5-8.5, however, the MDS projection reveals clear separation between groups of states (Fig. 5B), reflecting divergent mortality trajectories. States located in hotter and more climatically vulnerable regions form distinct clusters separated from cooler or high-altitude regions. This spatial reorganization of mortality behavior indicates that extreme warming drives structurally different heat-health responses across India.
The consistency between temporal divergence patterns (Fig. 4A and Fig. 4B) and spatial clustering behavior (Fig. 5A and Fig. 5B) reinforces the robustness of the projected regional differentiation in heat-related mortality risk.

\section{Discussions}
The growing divergence in projected heat-related mortality between the intermediate SSP2-4.5 and high-emission SSP5-8.5 scenarios underscores how emission choices will profoundly influence future urban heat-health risks. Under the intermediate mitigation scenario, mortality increases remain comparatively moderate and spatially uniform, whereas the high-emission pathway produces both a substantially higher overall mortality burden and increasing regional inequality in heat-related risk.
These findings are consistent with global assessments showing strong nonlinear amplification of extreme heat impacts under higher warming scenarios~\citep{IPCC_2021, Wang_2024}. As background temperatures rise, populations increasingly operate closer to physiological heat tolerance limits, making additional warming disproportionately more harmful. The accelerated mortality growth observed after mid-century under SSP5-8.5 reflects this compounding effect, emphasizing that delayed mitigation efforts can lock in long-term public health consequences.
Importantly, the contrast between scenarios demonstrates that climate mitigation is not only an environmental imperative but also a public health intervention. Lower-emission pathways substantially reduce the scale of future heat-related mortality and limit the emergence of extreme spatial disparities across urban regions.

The spatial patterns identified in this study indicate that future heat-related mortality risk is likely to evolve unevenly across India rather than increasing uniformly nationwide. Regions including the Deccan Plateau, western India, and parts of eastern and northeastern India exhibit more rapid escalation of projected mortality under the high-emission SSP5-8.5 pathway, whereas high-altitude and climatically cooler regions display comparatively slower growth.
These regional contrasts are consistent with previous findings linking baseline climate conditions, urban heat island intensity, and socioeconomic vulnerability to heat-related health outcomes~\citep{Mazdiyasni_2017, Srivastava_2022}. Cities located in hot semi-arid and tropical climatic zones already operate close to critical thermal stress thresholds, making them particularly sensitive to incremental warming. In addition, urban characteristics such as high building density, limited vegetation cover, extensive informal settlements, and constrained access to cooling infrastructure can further amplify heat exposure and reduce adaptive capacity.
The clustering behavior revealed by the multidimensional scaling (MDS) analysis suggests that extreme warming scenarios drive increasing structural divergence in regional mortality responses. This spatial reorganization reflects differences in climate sensitivity and exposure patterns across urban regions, highlighting that uniform national adaptation strategies may be insufficient. Instead, region-specific heat adaptation and resilience planning will be necessary to address emerging spatial inequalities in future heat-related health risk.

The projected rise in heat-related mortality across Indian cities has direct implications for urban planning and public health policy. Many Indian states and cities have already implemented Heat Action Plans (HAPs), focusing on early warning systems, public awareness campaigns, and emergency response mechanisms. However, the magnitude and spatial heterogeneity of future risk projected in this study suggest that existing interventions may be insufficient under high-emission futures.
Cities and regions identified as high-risk clusters under SSP5-8.5 should be prioritized for targeted adaptation investments. These include expansion of urban green infrastructure, reflective roofing and cool pavement technologies, improved building ventilation standards, and protection of vulnerable populations such as outdoor workers and elderly residents. Integrating heat risk projections into urban land-use planning and infrastructure design will be critical for reducing long-term exposure.
Furthermore, the relatively uniform mortality patterns observed under SSP2-4.5 emphasize the potential benefits of coordinated mitigation and adaptation strategies. Moderate warming pathways not only reduce absolute mortality burden but also limit regional inequality, supporting more equitable urban climate resilience outcomes.

Several limitations should be considered when interpreting the results of this study. First, climate projections were derived from global climate models with spatial resolutions of approximately 1°, which may underestimate localized temperature extremes and urban heat island intensity. As a result, fine-scale intra-urban variability in heat exposure is not fully resolved. Future studies incorporating high-resolution regional climate simulations and urban-scale downscaling frameworks could improve representation of localized thermal risk patterns.
Second, relative humidity and compound heat stress metrics such as wet-bulb temperature and apparent temperature were not explicitly incorporated due to incomplete availability across all CMIP6 models used in the ensemble. This limitation may lead to conservative estimates of heat-related mortality, particularly in humid coastal and monsoon-influenced regions where moisture strongly amplifies physiological heat stress \citep{Russo_2017, DiNapoli_2018}. Incorporation of multi-variable heat stress indices would provide a more physiologically representative assessment of extreme heat exposure in future projections.
Third, the ARIMAX modeling framework assumes an approximately linear relationship between temperature and mortality at annual timescales. Although this assumption is supported by previous epidemiological studies examining long-term temperature–mortality associations~\citep{Masselot_2022, Li_2025, Hasan_2025}, future research could explore nonlinear exposure–response relationships, threshold-based models, and lagged physiological effects to better capture short-term heat–health dynamics.
Finally, mortality data were available primarily at the national scale, limiting sub-state and intra-urban granularity and preventing direct assessment of socioeconomic heterogeneity in heat vulnerability. Integration of district-level mortality records, demographic projections, and vulnerability indicators such as age structure, housing quality, and access to cooling infrastructure would enable more detailed urban risk mapping and support more targeted adaptation planning in future analyses. Despite these limitations, several aspects of the projected mortality patterns remain robust. The strong divergence between SSP2-4.5 and SSP5-8.5 scenarios is consistently observed across the CMIP6 ensemble and reflects fundamental differences in long-term warming trajectories. While absolute mortality estimates may vary depending on model structure, climate sensitivity, and assumptions regarding future vulnerability, the relative amplification of heat-related mortality under high-emission pathways and the emergence of spatial inequality across regions represent stable features of the projections. These results therefore provide a robust indication of the potential public health consequences associated with different climate mitigation pathways.

\section{Conclusion}

This study investigates future heatwave-related mortality trends across Indian urban areas, integrating historical mortality data with CMIP6 temperature projections under moderate (SSP2-4.5) and high-emission (SSP5-8.5) scenarios. The results indicate a substantial increase in heat-related mortality throughout the 21st century, with markedly steeper growth under the high-emission scenario compared to the intermediate mitigation pathway. Beyond national trends, the analysis reveals a pronounced spatial variation in future heat-health risk. Under high-emission conditions, regional disparities intensify, with hotter and climatically vulnerable regions experiencing disproportionately higher mortality burdens. Clustering patterns identified through multidimensional scaling further demonstrate that extreme warming produces structurally different regional responses, reinforcing the importance of location-specific adaptation strategies.

These results emphasize that climate mitigation pathways play a critical role in shaping future heat-health risks. Limiting warming to intermediate scenarios such as SSP2-4.5 could substantially reduce both the magnitude and spatial inequality of heat-related mortality across Indian urban regions.

The divergence observed between the moderate SSP2-4.5 and high-emission SSP5-8.5 scenarios illustrates why both mitigation efforts and urban adaptation are essential. Lower emissions greatly decrease the aggregate mortality impact and promote greater equity, while high emissions threaten to establish extreme heat as a persistent, unevenly distributed health emergency. These findings underscore the urgency of integrating long-term heat risk projections into urban planning, public health preparedness, and climate-resilient infrastructure design. By connecting projected climate scenarios to urban mortality impacts at national and state levels, this work establishes an empirical basis for foreseeing heat-related health risks and informing targeted policies in a region facing among the world's highest levels of heat vulnerability.


\section*{Declaration of competing interest}
The authors declare that they have no known competing financial interests or personal relationships that could have appeared to influence the work reported in this paper.

\section*{Declaration of generative AI and AI-assisted technologies in the writing process}
During the preparation of this manuscript, the authors used ChatGPT (OpenAI) for language refinement and editing support. After using this tool, the authors reviewed and edited the content as needed and took full responsibility for the scientific accuracy, interpretation, and originality of the work.

\section*{Data availability}
CMIP6 climate projection data used in this study were obtained from the Copernicus Climate Data Store (CDS) under the ScenarioMIP and historical datasets (https://cds.climate.copernicus.eu/datasets/projections-cmip6). Observational reference temperature data from the India Meteorological Department (IMD) were accessed via the dataset reported in Mazdiyasni et al. (2017) (https://www.science.org/doi/10.1126/sciadv.1700066). Heat-related mortality data were compiled from publicly available reports and records published by the National Disaster Management Authority (NDMA), Ministry of Statistics and Programme Implementation (MoSPI), Ministry of Earth Sciences (MoES), and the National Crime Records Bureau (NCRB). The processed datasets and derived model outputs generated during the current study are available from the corresponding author upon reasonable request.

\section*{Acknowledgements}
Ingita Dey Munshi acknowledges the Department of Biotechnology (DBT), Government of India, for financial support in the form of a postgraduate fellowship. 

\bibliographystyle{elsarticle-num}
\bibliography{references}



\section*{Supplementary material}

# Supplementary Material for: Heatwave-Related Mortality Across Indian Cities Under Future Climate Scenarios

**S1. CMIP6 Climate Model Selection and Ensemble Construction**

Daily near-surface air temperature (T2M) and daily maximum temperature (Tmax) data were obtained from six CMIP6 global climate models: CNRM-CM6-1, CNRM-ESM2-1, EC-Earth3-CC, MIROC-ES2L, MPI-ESM1-2-LR, and MRI-ESM2-0 [1]. These models were selected based on (i) continuous availability of historical and SSP2-4.5 and SSP5-8.5 simulations, (ii) ability to represent large-scale circulation dynamics and land–atmosphere coupling processes relevant to the Indian monsoon and tropical climate system, and (iii) spatial coverage consistency over the Indian domain [2, 3, 4]

All datasets were standardized to a no-leap calendar format by removing February 29 from leap years to ensure temporal alignment across models. Temperature variables originally provided in Kelvin were converted to degrees Celsius prior to analysis.

An ensemble mean was constructed by averaging temperature fields across all six models at each time step. This approach reduces sensitivity to individual model biases and emphasizes robust large-scale climate signals, consistent with best practices in climate impact assessment.

*S1. Selected CMIP6 models and native horizontal resolution*

| Model Name | Country | Native Horizontal Resolution |
| --- | --- | --- |
| CNRM-CM6-1 | France | Atmosphere and land: ∼1.4° (∼140–150 km); Ocean and sea ice: ∼1° (∼100 km) |
| CNRM-ESM2-1 | France | Atmosphere and land: ∼1.4° (∼140–150 km); Ocean and sea ice: ∼1° (∼100 km) |
| EC-Earth3-CC | Europe (consortium) | Atmosphere and land: ∼1° (∼100 km); Ocean (NEMO ORCA1 grid): ∼1° (∼100 km, refined to 1/3° in tropics); Sea ice (LIM3): ORCA1 grid |
| MIROC-ES2L | Japan | Atmosphere: ∼2.8° (∼280 km, T42 spectral truncation); Ocean: ∼1° longitude × 0.5–1° latitude |
| MPI-ESM1-2-LR | Germany | Atmosphere: ∼200 km (T63 spectral truncation); Ocean: ∼1.5° |

| Model Name | Country | Native Horizontal Resolution |
|---|---|---|
| MRI-ESM2-0 | Japan | Atmosphere: $\sim 1.1° \times 1.1°$ ($\sim$120 km); Ocean and sea ice: $\sim 1°$ zonal, 0.3–0.5° meridional (30–50 km in Arctic) |

## S2. Urban Location Selection and Geographic Coverage

A total of 67 locations were selected across India to represent diverse climatic zones, urban typologies, and geographic regions. These locations include coastal cities, inland plains, plateau regions, semi-arid environments, and high-altitude settlements. Selection was guided by the spatial resolution of CMIP6 models (approximately 1° grid spacing) to ensure that each location corresponded closely to a representative climate grid cell.

The spatial distribution ensures coverage across major climatic regimes, including the Indo-Gangetic Plain, Western Ghats, Deccan Plateau, northeastern hill regions, and coastal belts.

*Table S2. Geographic coordinates of selected locations*

| S.No. | Location | Latitude | Longitude | Urban/Rural |
|---|---|---|---|---|
| 1 | Agartala | 23.83 | 91.28 | Urban |
| 2 | Aizawl | 23.73 | 92.72 | Urban |
| 3 | Akola | 20.71 | 77.00 | Urban |
| 4 | Amaravati | 16.50 | 80.65 | Urban |
| 5 | Anantapur | 14.68 | 77.60 | Rural |
| 6 | Aurangabad | 19.88 | 75.32 | Urban |
| 7 | Bareilly | 28.36 | 79.41 | Urban |
| 8 | Baripada | 21.94 | 86.72 | Urban (Odisha - rural-dominated state) |
| 9 | Bengaluru | 12.97 | 77.59 | Urban |
| 10 | Bhopal | 23.25 | 77.41 | Urban |
| 11 | Bhubaneswar | 20.27 | 85.84 | Urban (Odisha - rural-dominated state) |
| 12 | Bhuj | 23.25 | 69.67 | Urban |
| 13 | Bikaner | 28.01 | 73.31 | Rural |
| 14 | Chandigarh | 30.73 | 76.78 | Urban |
| 15 | Chandrapur | 19.95 | 79.30 | Rural |
| 16 | Chennai | 13.08 | 80.27 | Urban |
| 17 | Daman | 20.42 | 72.85 | Urban |
| 18 | Dehradun | 30.32 | 78.03 | Urban |
| 19 | Delhi | 28.61 | 77.23 | Urban |
| 20 | Dispur | 26.14 | 91.77 | Urban |
| 21 | Durgapur | 23.55 | 87.32 | Urban |
| 22 | Gandhinagar | 23.22 | 72.65 | Urban |
| 23 | Gangtok | 27.33 | 88.61 | Urban (Sikkim - rural-dominated state) |
| 24 | Gorakhpur | 26.76 | 83.37 | Urban |

| | | | | |
|---|---|---|---|---|
| 25 | Gulbarga | 17.33 | 76.83 | Urban |
| 26 | Gwalior | 26.22 | 78.18 | Urban |
| 27 | Hubli-Dharwad | 15.36 | 75.12 | Urban |
| 28 | Hyderabad | 17.39 | 78.49 | Urban |
| 29 | Imphal | 24.82 | 93.95 | Urban |
| 30 | Itanagar | 27.10 | 93.62 | Urban |
| 31 | Jabalpur | 23.18 | 79.98 | Urban |
| 32 | Jaipur | 26.91 | 75.79 | Urban |
| 33 | Jaisalmer | 26.91 | 70.91 | Rural |
| 34 | Jodhpur | 26.24 | 73.02 | Urban |
| 35 | Kavaratti | 10.57 | 72.64 | Urban |
| 36 | Kohima | 25.67 | 94.11 | Rural |
| 37 | Kolkata | 22.57 | 88.36 | Urban |
| 38 | Koraput | 18.81 | 82.71 | Rural |
| 39 | Kota | 25.18 | 75.83 | Urban |
| 40 | Leh | 34.15 | 77.57 | Urban |
| 41 | Lucknow | 26.85 | 80.95 | Urban |
| 42 | Mahbubnagar | 16.74 | 77.98 | Rural |
| 43 | Mangalore | 12.91 | 74.85 | Urban |
| 44 | Mumbai | 19.08 | 72.88 | Urban |
| 45 | Munnar | 10.09 | 77.06 | Rural |
| 46 | Mysore | 12.30 | 76.65 | Urban |
| 47 | Nagpur | 21.15 | 79.09 | Urban |
| 48 | Palamu | 24.08 | 84.08 | Rural |
| 49 | Panaji | 15.49 | 73.83 | Urban |
| 50 | Pasighat | 28.07 | 95.33 | Urban |
| 51 | Patna | 25.61 | 85.15 | Urban |
| 52 | Port Blair | 11.62 | 92.72 | Urban |
| 53 | Puducherry | 11.93 | 79.83 | Urban |
| 54 | Pune | 18.52 | 73.85 | Urban |
| 55 | Raipur | 21.25 | 81.63 | Urban |
| 56 | Rajkot | 22.30 | 70.78 | Urban |
| 57 | Ranchi | 23.35 | 85.33 | Urban |
| 58 | Sambalpur | 21.47 | 83.98 | Urban |
| 59 | Shillong | 25.57 | 91.88 | Urban |
| 60 | Shimla | 31.10 | 77.16 | Urban |
| 61 | Srinagar | 34.08 | 74.80 | Urban |
| 62 | Surat | 21.17 | 72.83 | Urban |
| 63 | Thiruvananthapuram | 8.52 | 76.93 | Urban + Rural |
| 64 | Tiruchirapalli | 10.79 | 78.70 | Urban |
| 65 | Udaipur | 24.58 | 73.68 | Urban + Rural |
| 66 | Ujjain | 23.18 | 75.78 | Urban + Rural |
| 67 | Varanasi | 25.32 | 82.98 | Urban + Rural |

## S3. Climate Model Temperature Processing and Baseline Alignment

To define an appropriate seasonal window for computing Summer Mean Temperature (SMT), an exploratory analysis was conducted using historical daily maximum temperature (Tmax) data from 1970 to 2014 across all 67 selected locations. For each location and year, the hottest day of the year was identified, and the corresponding calendar month was recorded. The resulting distribution of hottest-day months across all locations and years is shown in Fig. S1.

The analysis indicates that peak annual temperature extremes in India predominantly occur between March and September, with the highest concentration during April, May, and June. However, a non-negligible fraction of extreme temperature events also occurs during the transition periods of the pre-monsoon and early monsoon seasons, particularly in March and July–August for several regions.

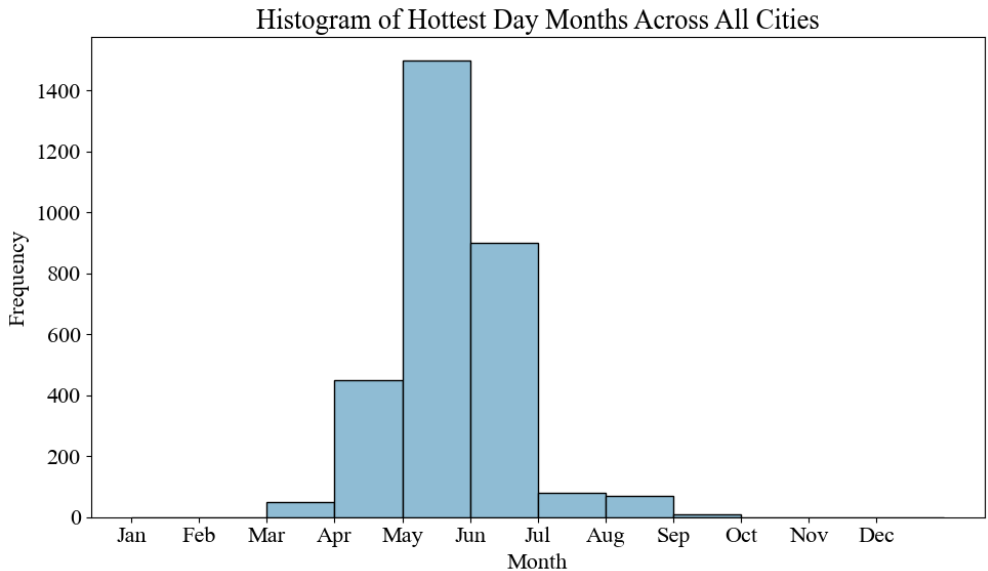

Fig. S1: Distribution of months corresponding to the hottest day of the year across 67 locations during the historical period (1970–2014). The histogram shows that peak annual temperature extremes predominantly occur between March and September, supporting the selection of this period for computing Summer Mean Temperature (SMT).

Additional exploratory analysis using location-specific violin distributions of annual Tmax peaks (not shown) confirmed consistent clustering of extreme temperature occurrences during the late pre-monsoon and early summer months across diverse climatic zones. Based on this combined evidence, the SMT period was defined as March to September to ensure inclusion of the full seasonal window during which extreme heat conditions occur across India [5]. This extended definition avoids truncation of heat exposure signals that may result from restricting analysis to shorter seasonal windows.

To ensure consistency between climate model outputs and observed temperature magnitudes, a baseline alignment procedure was applied using ground-based observations from the India Meteorological Department (IMD). Summer Mean Temperature (SMT) was calculated for both observed and modeled datasets over the overlapping reference period (1970–2006).

A mean offset adjustment approach was implemented by computing the average difference between modeled and observed SMT during the reference period and applying this adjustment factor to projected SMT values for future periods. Observed IMD values were retained directly for the historical period, while adjusted CMIP6 ensemble SMT values were used for projection analysis.

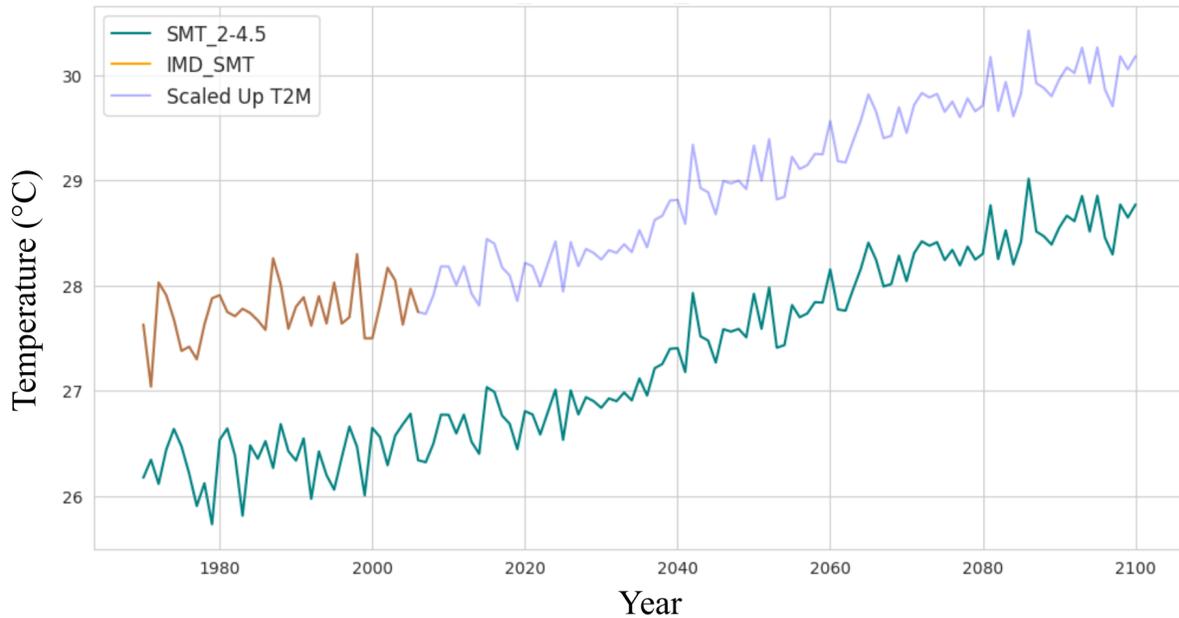

Fig. S2: Model–observation baseline alignment of ensemble Summer Mean Temperature (SMT). The figure compares observed IMD SMT, raw CMIP6 ensemble SMT, and baseline-adjusted SMT used for projection analysis.

This approach preserves long-term warming trends from climate models while aligning baseline temperature levels with observed climatology. Fig. S2 illustrates the comparison between raw model output, observed temperature, and baseline-adjusted SMT time series.

### S4. Heatwave Definition Sensitivity Assessment and Threshold Selection

Daily maximum temperature (Tmax) data were processed to identify heatwave days using three alternative definitions to assess the sensitivity of heatwave metrics to threshold choice and event duration criteria:

1. **95th Percentile (3+ Days)** – Heatwave events defined as three or more consecutive days during which Tmax exceeded the 95th percentile of the historical temperature distribution for each location [6].
2. **95th Percentile (2+ Days)** – Heatwave events defined as two or more consecutive days exceeding the 95th percentile threshold [7].
3. **IMD Operational Definition** – Heatwave events defined according to the India Meteorological Department (IMD) criteria, where Tmax exceeds absolute temperature

5

thresholds of ≥40°C for plains, ≥37°C for coastal regions, and ≥30°C for hilly regions [8].

For percentile-based approaches, the 95th percentile threshold was computed independently for each location using the historical Tmax distribution. This procedure follows commonly adopted climatological approaches for identifying relative temperature extremes [9]. An example of percentile threshold estimation based on historical Tmax distributions is illustrated in Fig. S3.

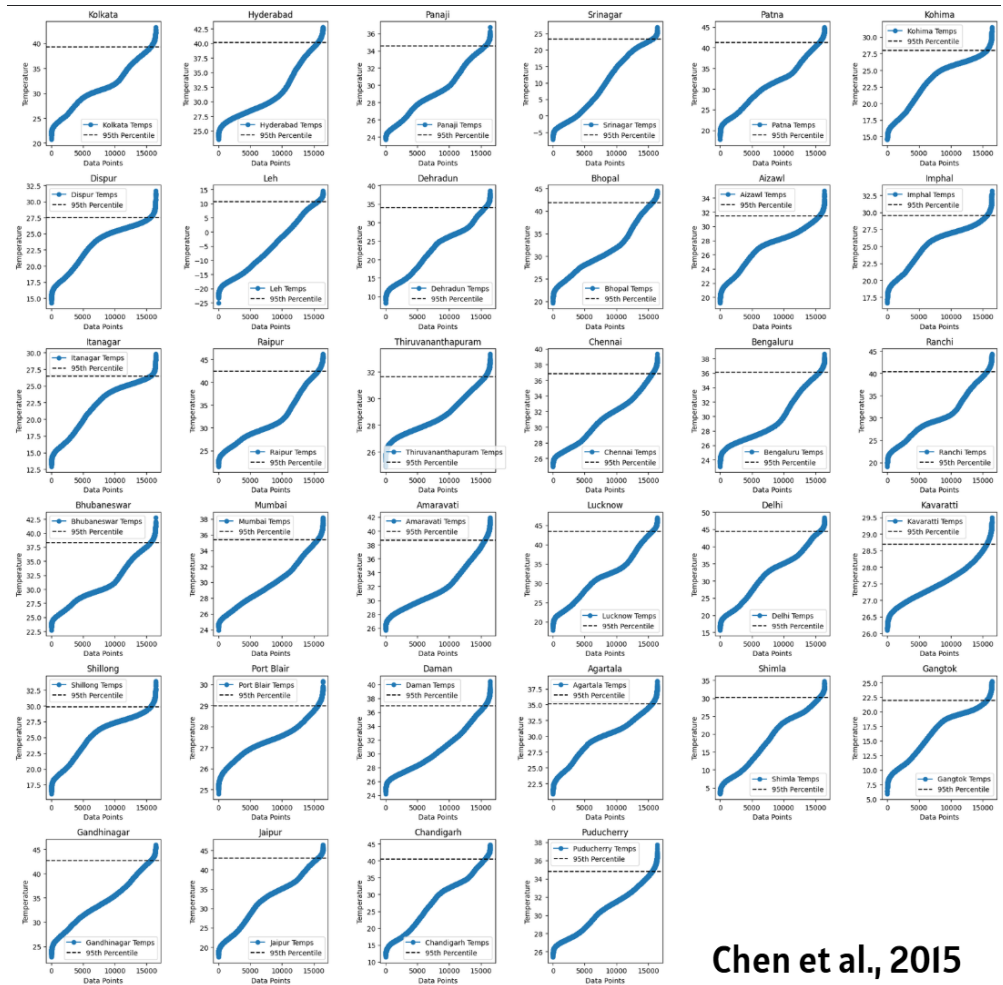

Fig. S3: Example illustration of 95th percentile Tmax threshold estimation based on historical daily temperature distributions for selected urban locations. The dashed horizontal line indicates the location-specific 95th percentile threshold used in percentile-based heatwave identification methods.

Although percentile-based definitions are useful for capturing relative extremes, their application across historical and future projection periods introduced methodological artifacts. Specifically, the 95th percentile thresholds for historical periods (centered near 2014) and scenario periods (centered near 2100) differed substantially due to climate-driven shifts in the underlying temperature distribution. This resulted in an artificial dip in national heatwave day counts at the historical-to-scenario transition boundary (2014–2015), as illustrated in Figure S4.

6

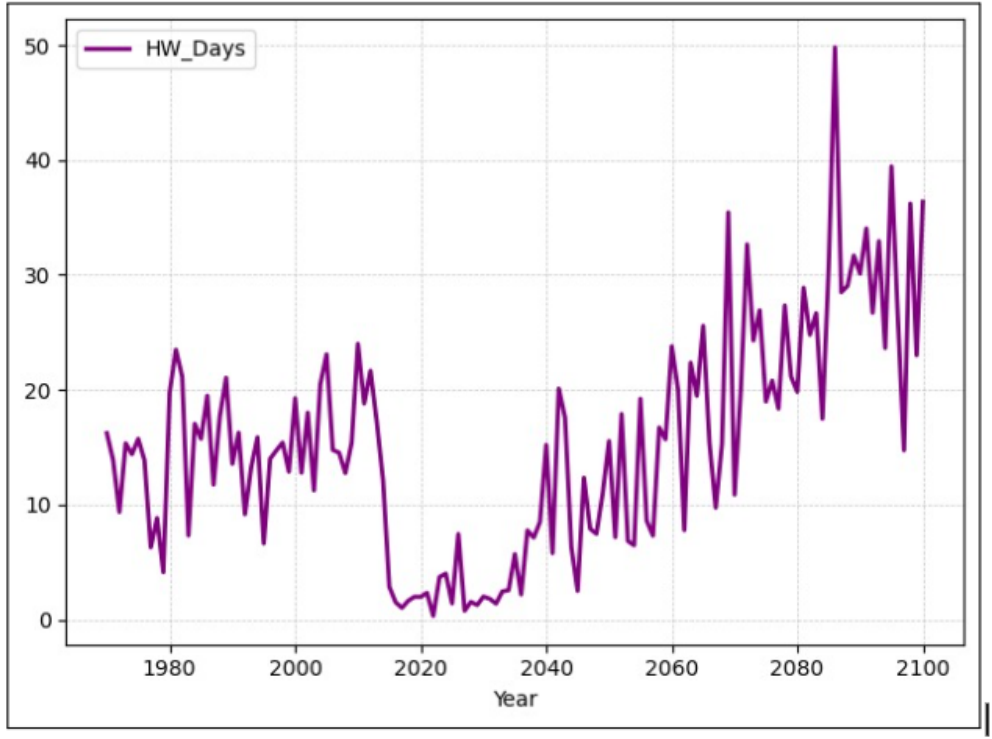

Fig. S4: National average heatwave days per year computed using 95th percentile–based definitions (1970–2100). The artificial dip observed during the historical-to-scenario transition period (2014–2015) results from shifting percentile baselines, where thresholds are computed separately for historical and future climate distributions. While using a single combined historical–future distribution could remove this discontinuity, such an approach would artificially inflate early-period heatwave counts and distort the representation of relative extremes. To ensure physical interpretability and temporal consistency, the IMD absolute threshold-based heatwave definition was adopted for the primary analysis.

An alternative approach of computing percentile thresholds using a combined historical and future temperature distribution was also evaluated. However, this method resulted in unrealistically high heatwave counts during early historical years due to upward shifts in the overall temperature distribution driven by future warming. Such inflation of early-period heatwave frequency reduces physical interpretability and biases trend assessment. Therefore, percentile thresholds were computed separately for historical and scenario periods for sensitivity testing only and were not used in the primary analysis.

To ensure temporal continuity, operational relevance, and consistency across historical and future periods, the IMD threshold-based definition [8] was adopted for the primary heatwave analysis. Unlike percentile-based thresholds, the IMD definition remains fixed in absolute temperature space and avoids discontinuities arising from shifting climatological baselines. This approach also aligns with nationally recognized operational heatwave monitoring practices used for public health advisories and disaster management in India.

Annual national heatwave day counts were computed by applying the IMD definition independently at each of the 67 locations, summing annual heatwave days across all locations, and dividing by the total number of locations to obtain a spatially averaged national heatwave metric.

## S5. Selection of Summer Mean Temperature as the Primary Climate Predictor

To evaluate the suitability of Summer Mean Temperature (SMT) as the primary climate predictor in the mortality forecasting framework, comparative analyses were conducted between heatwave day counts derived using the IMD definition and SMT time series [5].

First, both variables were standardized using z-score normalization to enable direct comparison of long-term temporal behavior. The standardized time series comparison (Figure S5) demonstrates strong temporal coherence between CMIP6-derived heatwave day counts and scaled SMT across historical and future periods. Both variables exhibit consistent long-term upward trends and similar interannual variability patterns, indicating that SMT captures the dominant climate signal associated with increasing heatwave activity.

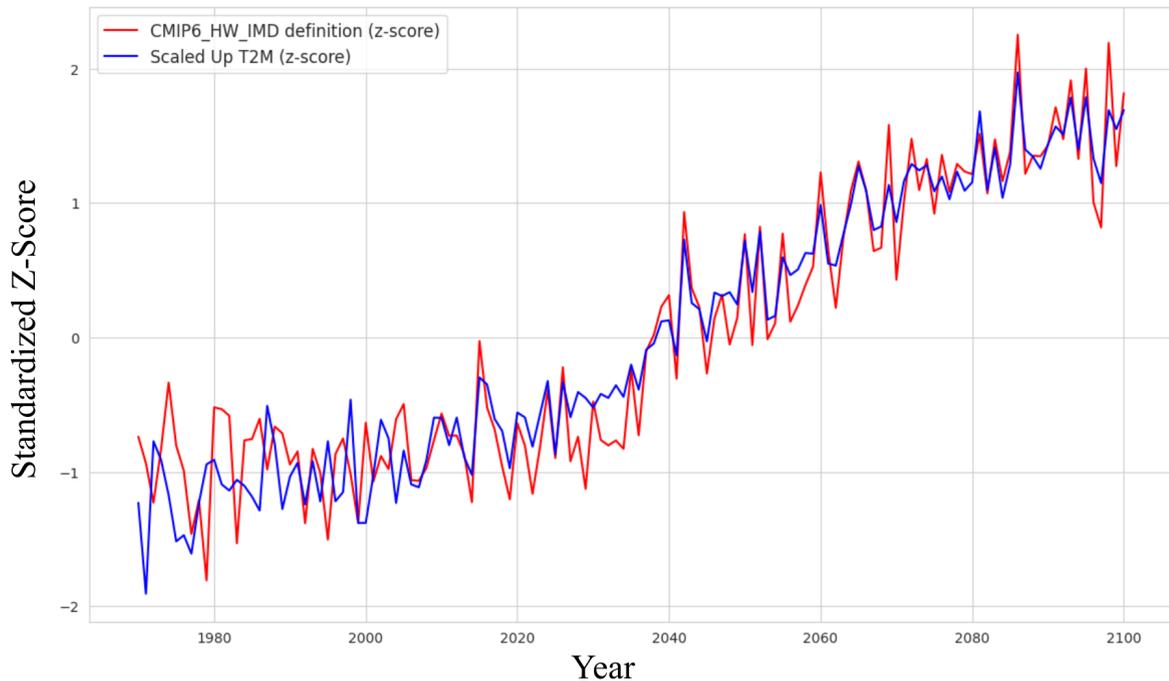

Fig. S5: Standardized (z-score) time series comparison between CMIP6-derived heatwave day counts (IMD definition) and scaled Summer Mean Temperature (SMT) from 1970 to 2100. The strong temporal coherence demonstrates that SMT captures long-term trends and interannual variability associated with increasing heatwave activity across historical and future periods.

Second, the functional relationship between annual heatwave day counts and SMT was examined separately for SSP2-4.5 and SSP5-8.5 scenarios (Figure S6). Scatter analysis reveals a strong linear association between SMT and heatwave frequency under both emission pathways, with higher SMT values corresponding to substantially increased heatwave occurrence. This relationship remains stable across scenarios, demonstrating robustness of SMT as a proxy for cumulative seasonal heat stress.

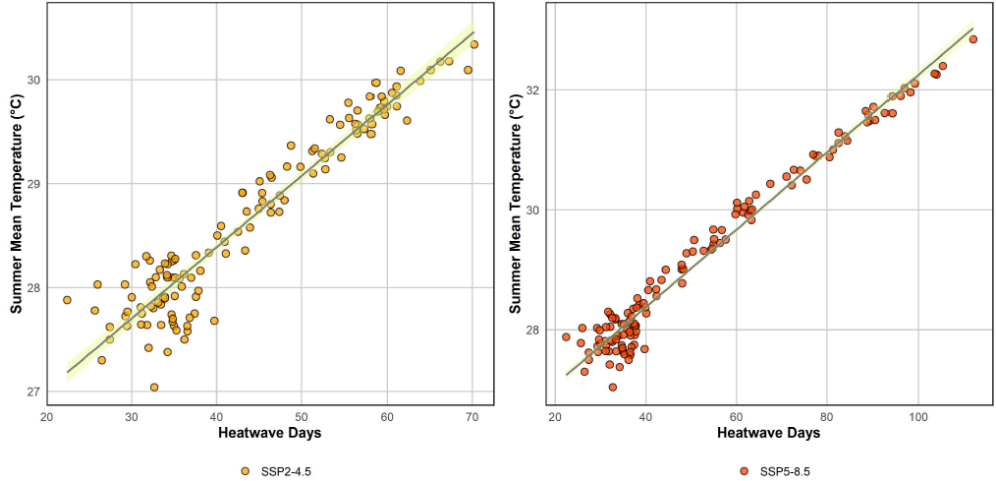

Fig. S6: Relationship between annual heatwave day counts and Summer Mean Temperature (SMT) under SSP2-4.5 (left) and SSP5-8.5 (right) scenarios. Scatter plots with fitted regression lines illustrate strong linear associations between cumulative summer warming and heatwave frequency across emission pathways.

Including both SMT and heatwave day counts as predictors introduced strong multicollinearity due to their high correlation. Given its continuous nature, smoother temporal behavior, and consistent availability across CMIP6 models and scenarios, SMT was selected as the primary exogenous climate driver for ARIMAX modeling [10]. Heatwave day metrics were retained for descriptive analysis and physical interpretation but were not directly incorporated into the forecasting model to preserve numerical stability and interpretability.

## S6. Mortality Dataset Processing and Quality Control

National heatwave-related mortality data were compiled from official government sources, including the India Meteorological Department (IMD), National Disaster Management Authority (NDMA), Ministry of Statistics and Programme Implementation (MoSPI), Ministry of Earth Sciences (MoES), and the National Crime Records Bureau (NCRB). These sources provide consolidated annual heatstroke-related fatality records at the national level.

When overlapping records were available for the same year across multiple official sources, an arithmetic mean was computed to derive a consolidated national mortality estimate. This ensemble-averaging approach reduces sensitivity to reporting inconsistencies across agencies and provides a more stable representation of annual mortality variability. Additional quality control procedures included cross-checking year-to-year continuity, removal of duplicate entries, verification of reporting anomalies, and consistency checks across multiple data sources.

No spatial aggregation was required because mortality values were directly reported at the national scale. The final dataset spans 1970–2023 and represents the dependent variable used for ARIMAX model training and validation. This time series captures both long-term mortality trends and interannual variability associated with extreme heat exposure.

## S7. ARIMAX Model Configuration and Validation Strategy

Future heat-related mortality projections were generated using an Autoregressive Integrated Moving Average model with exogenous regressors (ARIMAX), a widely used frame-

9

work for modeling time-series processes with external predictors [11]. Since mortality data were available at the yearly level, no seasonal periodicity was present and a non-seasonal specification was therefore adopted.

An ARIMAX$(1, 0, 1)$ model was selected following the Box–Jenkins time-series modeling framework, which identifies appropriate autoregressive and moving-average orders through inspection of autocorrelation structure and model diagnostics [11].

In backshift operator notation, the model can be expressed as:

$$(1 - \phi_1 B)Y_t = (1 + \theta_1 B)\varepsilon_t + \beta\, SMT_t \tag{1}$$

where $B$ denotes the backshift operator ($BY_t = Y_{t-1}$), $Y_t$ represents annual heat-related mortality, $SMT_t$ is Summer Mean Temperature, $\phi_1$ is the autoregressive parameter capturing temporal persistence in mortality, $\theta_1$ is the moving-average parameter accounting for serially correlated shocks, $\beta$ measures sensitivity of mortality to summer mean temperature, and $\varepsilon_t$ is assumed to be white noise with zero mean and constant variance [11].

Prior to model estimation, both mortality and Summer Mean Temperature (SMT) series were standardized over the historical calibration period to ensure comparable scaling and stable parameter estimation. Model parameters were estimated using annual data from 1970 to 2023.

Since the primary aim of the framework is to compare long-term climate scenarios rather than generate short-term forecasts, emphasis was placed on capturing consistent trend behavior and a stable temperature–mortality relationship rather than optimizing conventional short-term forecast accuracy metrics such as RMSE or MAPE.

## S8. State-Level Climate Exposure Aggregation and Temporal Smoothing

Projected city-level temperature time series were spatially assigned to their corresponding administrative states and union territories. For each state, Summer Mean Temperature (SMT) values were averaged across all associated urban locations to generate representative regional climate exposure time series.

State-level mortality trajectories were generated by applying the trained national ARIMAX model to state-specific SMT time series. While the ARIMAX model was calibrated using national mortality and national SMT, regional SMT inputs were used during the forecasting stage to capture spatial variability in climate-driven mortality response. This approach enables projection of region-specific mortality trajectories driven by spatial heterogeneity in climate exposure while maintaining a common epidemiological response structure learned from national-scale mortality data.

To enhance visualization and reduce short-term interannual variability, projected mortality trajectories were temporally smoothed using an eight-year moving average with a one-year step. This smoothing procedure was applied exclusively for graphical presentation and spatial pattern analysis, including the generation of state-level spatial maps (Fig. S7) and multidimensional scaling (MDS) visualizations. Smoothed values were not used during ARIMAX model training or mortality projection computations.

To provide a continuous visualization of the temporal evolution of state-level projected mortality patterns from 2024 to 2100 under both SSP2-4.5 and SSP5-8.5 scenarios, an animated supplementary visualization is provided as Supplementary Video S1 (available at:

https://drive.google.com/file/d/1WyFnnCOIgec-jABBVG-VADh2DQdRgbFu/view?usp=sharing). The animation illustrates the progressive amplification of regional disparities and the emergence of spatial inequality under high-emission conditions.

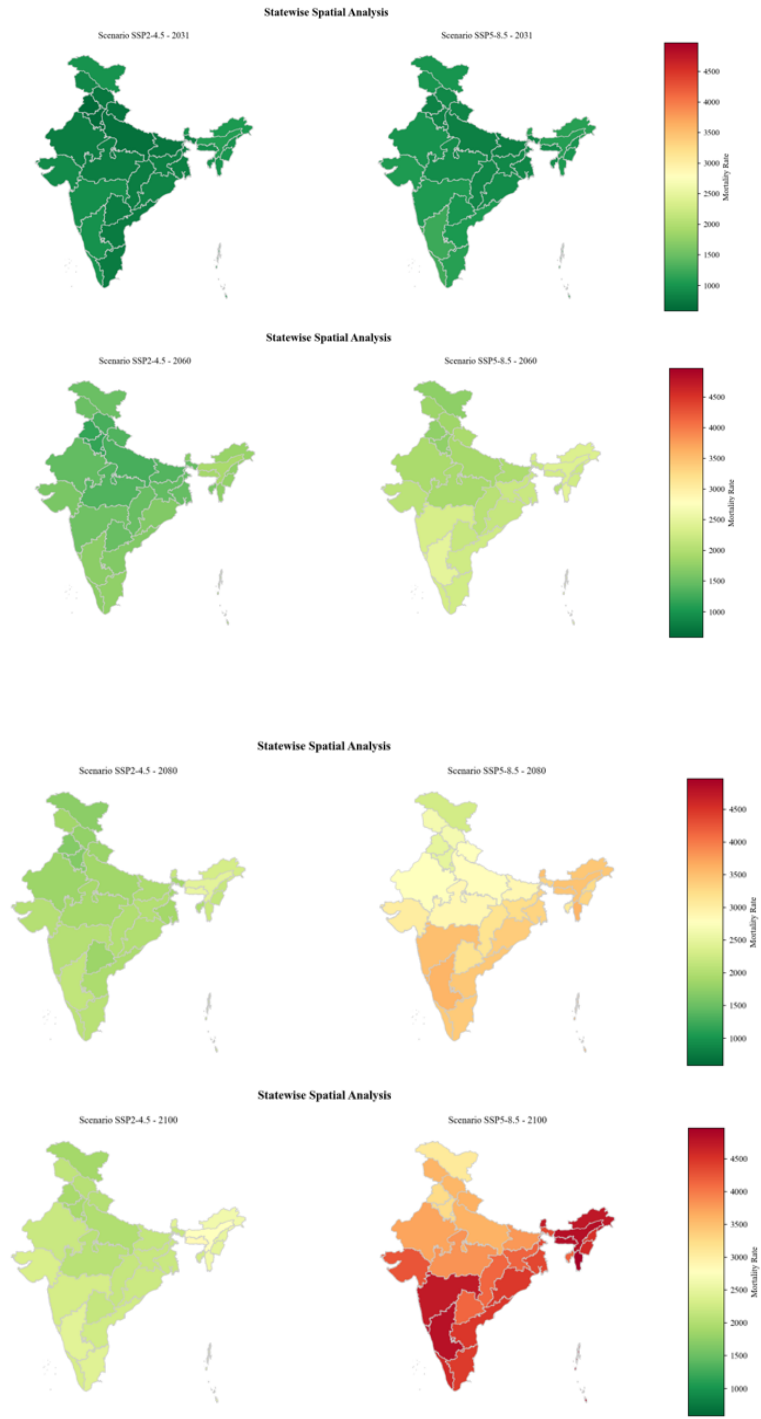

Fig. S7: Spatial distribution of projected relative heat-related mortality across Indian states and union territories derived by forcing the nationally trained ARIMAX model with state-aggregated SMT projections under SSP2-4.5 and SSP5-8.5 scenarios for selected future periods (2031, 2060, 2080, and 2100). Maps illustrate progressive amplification of mortality burden and increasing spatial heterogeneity under high-emission pathways.

Spatial patterns emerging from the state-level projections indicate increasing vulnerability in regions such as the Deccan Plateau and parts of northeastern India under future warming scenarios (Fig. S7). The Deccan region exhibits strong amplification due to persistently high baseline temperatures combined with projected intensification of summer heat, while northeastern states show increasing sensitivity driven by warming trends in historically moderate climate zones. These patterns reflect the combined influence of baseline climatology, regional warming rates, and expanding exposure of populations to extreme heat conditions, providing physical context for the spatial divergence patterns examined in subsequent clustering analyses.

## S9. Multidimensional Scaling (MDS) Distance Metric and Implementation

Multidimensional scaling (MDS) is a dimensionality reduction technique that represents high-dimensional data in a low-dimensional space while preserving pairwise dissimilarity relationships as faithfully as possible [12, 13]. In the context of this study, each state is characterized by a long projected mortality time series, resulting in a high-dimensional representation that is difficult to interpret visually. The mortality trajectories used for MDS analysis correspond to state-specific projections generated by driving the trained national ARIMAX model with regional SMT inputs. MDS was therefore used to project these temporal trajectories into a two-dimensional space, enabling visualization and interpretation of similarity and divergence patterns in state-level heat-related mortality behavior.

MDS was selected for this analysis because the primary objective was not classification but exploratory assessment of structural similarity among long-term mortality trajectories. Unlike clustering algorithms that impose discrete group boundaries, MDS preserves continuous similarity relationships and allows identification of gradual transitions and emerging spatial divergence patterns under different emission scenarios [13].

*Distance Metric Selection*

To quantify pairwise similarity between state-level mortality trajectories, a correlation-based distance metric derived from Pearson correlation was employed. This metric emphasizes similarity in temporal evolution patterns rather than absolute magnitude differences, which is particularly important when comparing regions with different baseline mortality levels but potentially similar climate-driven response behavior.

For any two states $i$ and $j$, the pairwise distance $d_{ij}$ was computed as:

$$d_{ij} = \sqrt{2(1 - C_{ij})} \tag{2}$$

where $C_{ij}$ denotes the Pearson correlation coefficient between the projected mortality time series of states $i$ and $j$ computed over the full projection period. This transformation converts correlation similarity into a Euclidean distance metric suitable for multidimensional scaling while preserving the relative similarity structure of the data [14].

To ensure hierarchical consistency in similarity relationships, the resulting distance matrix was evaluated against the ultrametric inequality

$$d(x, z) \leq \max\{d(x, y), d(y, z)\} \tag{3}$$

13

which enforces a strong triangle inequality characteristic of ultrametric spaces. This condition ensures that similarity relationships among states reflect hierarchical structure in temporal mortality trajectories.

The resulting distance matrix was provided as input to classical multidimensional scaling (MDS) to generate low-dimensional representations of state-level mortality trajectories [12]. MDS preserves continuous similarity relationships without imposing discrete cluster boundaries, enabling exploratory assessment of spatial reorganization and gradual divergence patterns under different emission scenarios.

Both two-dimensional and three-dimensional embeddings were evaluated during exploratory analysis. While the three-dimensional representation retained additional structural detail, the two-dimensional configuration was selected for presentation because it provided clearer visual separation and interpretability of clustering patterns relevant to regional similarity assessment.

The resulting MDS projections were interpreted as relative similarity maps of long-term mortality response behavior, where spatial proximity in the embedded space reflects similarity in temporal evolution rather than geographic distance.

### S10. Additional Uncertainty Sources and Interpretation Considerations

Uncertainty in projected mortality outcomes arises from multiple sources, including inter-model variability within the CMIP6 ensemble, emission pathway assumptions, temperature baseline alignment procedures, and variability in heat-related mortality reporting practices. While ensemble averaging reduces sensitivity to individual climate model behavior, absolute mortality projections should be interpreted as indicative trajectories rather than precise forecasts.

The present analysis focuses on climate-driven mortality sensitivity and does not explicitly incorporate future population growth, demographic transitions, urban expansion, or adaptive responses such as cooling infrastructure deployment and heat action plans. These factors may substantially influence real-world health outcomes and should be incorporated in future integrated assessment frameworks that combine climate, demographic, and urban development pathways. Additional uncertainty arises from the assumption of spatially uniform temperature–mortality sensitivity when applying the national ARIMAX model to regional SMT inputs.

Despite these uncertainties, relative differences between emission pathways and spatial divergence patterns across regions remain robust across modeling steps. The results therefore provide reliable comparative insights into the magnitude and spatial structure of future heat-related mortality risk, supporting the primary conclusions regarding the importance of mitigation and region-specific urban heat adaptation planning.



\end{document}